\documentclass[11pt,a4paper]{article}
\usepackage{amsmath}
\usepackage[round,authoryear]{natbib}
\usepackage{graphics}
\usepackage{amsfonts}
\usepackage{amssymb}%
\usepackage{geometry}
\usepackage{graphicx}% Include figure files
\usepackage{dcolumn}
\usepackage{setspace}
\usepackage{booktabs}
\usepackage{array}
\usepackage[table]{xcolor}
\usepackage[colorlinks=true, linkcolor=my-magenta, citecolor=light-sky, urlcolor=my-magenta]{hyperref}

\usepackage{algorithm}
\usepackage{algpseudocode} % optional, if you want pseudocode commands

\usepackage{caption}
\usepackage{siunitx}
\usepackage{enumitem}  % allows custom enumerate labels
\providecommand{\U}[1]{\protect\rule{.1in}{.1in}}

\newtheorem{Ass'}{Assumption'}

\newcommand{\be}{\begin{equation}}
\newcommand{\ee}{\end{equation}}

\definecolor{light-sky}{RGB}{30, 120, 200}
\definecolor{sky}{RGB}{18,80,142}
\definecolor{dark-sky}{RGB}{18,90,150}
\definecolor{light-gray}{gray}{0.9}
\definecolor{bggray}{gray}{0.8}
\definecolor{olive}{RGB}{1, 69, 44}
\definecolor{aqua}{RGB}{127,219,255}
\definecolor{purple}{RGB}{122,0,153}
\definecolor{pink}{RGB}{255,0,153}
\definecolor{my-magenta}{rgb}{0.79, 0.08, 0.48}
\definecolor{my-magenta-light}{rgb}{0.85, 0.15, 0.55}
\definecolor{coolblack}{rgb}{0.0, 0.18, 0.39}
\definecolor{myBlue}{HTML}{1A73E8}

\DeclareMathOperator{\E}{E} % expectation

\begin{document}
\title{Benchmarking Formula~1 results using a normal model}

\author{
John Fry\thanks{Corresponding author. School of Business, University of Leicester, Leicester LE2 1RQ, UK.} \and
Silvio Fanzon\thanks{Department of Mathematics, School of Digital and Physical Sciences, University of Hull, Hull HU6 7RX, UK.\\ E-mail addresses: \href{mailto:frymaths@googlemail.com}{frymaths@googlemail.com} (J.~Fry), 
\href{mailto:S.Fanzon@hull.ac.uk}{S.Fanzon@hull.ac.uk} (S.~Fanzon),
\href{mailto:m.austin-2020@hull.ac.uk}{m.austin-2020@hull.ac.uk} (M.~Austin),
\href{mailto:thomasbrighton02@gmail.com}{thomasbrighton02@gmail.com} (T.~Brighton)}
\and
Mark Austin\footnotemark[2]
\and
Tom Brighton\footnotemark[2]
}

\date{}

\maketitle

\begin{abstract}
\noindent There is enduring interest in disentangling the effects of skill and luck in sport. A key issue in Formula 1 is distinguishing between car-level and driver-level effects. Four elite teams currently dominate Formula 1 and have won every major race for the last four years. In this paper we use univariate and bivariate normal models to quantify reasonable performance expectations at both driver and team levels, distinguishing between elite and non-elite teams. We illustrate our approach with an application to the last fully completed 2025 season. \\[0.5em]  
\end{abstract}

\section{Introduction} \label{sec:introduction}

Key questions in sports analysis concern what constitutes reasonable performance expectations \citep{FryJBR2021} and how to distinguish the effects of luck and skill \citep{Scarf2}. Though often under-explored, Formula~1 and motorsport provide a rich source of economic problems \citep{Wesselbaum}, particularly in light of the sport's rapid recent expansion. This growth has been supported by increased media and digital engagement. This includes the reality-style Netflix series \textit{Drive to Survive}, regulatory reforms designed to enhance competition, and strategic expansion into key markets like the United States with additional races in Miami and Las Vegas. This has significantly increased the sport's economic scale and global reach.

Analysing performance in Formula 1 is complicated by the need to disentangle car-level and driver-level effects \citep{FryF1}. Teams may also pursue objectives beyond simply winning like improving on the previous season, outperforming close rivals, securing higher prize-money, and strengthening longer-term competitive standing \citep{MouraoBook}.

At the time of writing, Formula~1 consists of ten teams, each fielding two drivers. Four elite teams -- Red Bull, Mercedes, Ferrari, and McLaren -- dominate the sport and, at the time of writing, had won every major race over the past four seasons. Despite this dominance, competition within these confines remains intense. In this paper, we investigate reasonable performance expectations under these conditions, explicitly accounting for the distinction between elite and non-elite teams. Furthermore, within a team, driver finishing positions are negatively correlated, violating the independence assumption common to most sports' models \citep{Scarf}. Building on \cite{FryF1}, we develop a novel framework based on univariate and bivariate normal distributions that simultaneously addresses both the elite/non-elite distinction and within-team correlations. This provides a more realistic evaluation of performance, and captures strategic and competitive aspects of the sport. We obtain analytical expressions for winning probabilities and supplement these with Monte Carlo simulation analyses. These simulations establish benchmark expectations for elite and non-elite drivers and teams. This enables us to identify which drivers and teams exceed, meet, or fall short of reasonable performance expectations.

Our analysis reveals that elite drivers Norris, Piastri and Verstappen performed well above expectations. However, several other elite-team drivers seem to under-perform. Many non-elite drivers also exceeded expectations, highlighting the depth of talent the sport. It is important to emphasise that non-elite teams also include exceptional drivers, notably Fernando Alonso, who is widely regarded as one of the most talented drivers on the grid but is arguably constrained by a relatively modest Aston Martin car. Statistical evidence of Alonso outperforming his machinery is provided in \cite{FryF1}. At the team level, McLaren exceeded expectations, whilst other elite teams Red Bull, Ferrari and Mercedes under-performed. Results obtained thus cast the sport in a new light and offer a more nuanced understanding of performance across teams and drivers.

The layout of this paper is as follows. Section~\ref{sec:univariate} presents a univariate model for individual driver performance, while Section~\ref{sec:bivariate} extends this to a bivariate framework for team performance. Section~\ref{sec:empirical} provides empirical applications, including performance benchmarking of drivers and teams in the last fully completed 2025 Formula~1 season. Section~\ref{sec:conclusion} concludes. Additional benchmarking considerations are discussed in Appendix~\ref{appendix}.

\section{Univariate modelling of individual driver performance} \label{sec:univariate}

In this section we outline a univariate model to describe the performance of individual drivers. Teams are classified into elite and non-elite categories, with 8 drivers in elite-team cars (two drivers per team across 4 teams) and 12 drivers in non-elite cars (across 6 teams). We assume that successive races are independent. Let the final finishing position in a race be approximated by a normal distribution. Whilst this may seem a strong assumption, this is consistent with the use of regression-based models in applications \citep{FryF1}. For elite-team drivers, we set
\begin{align}
& \E\left[\mbox{Elite driver Finishing Position}\right]  = \E \left[ \operatorname{U}\{1, 8\} \right] = 4.5, \label{eq1}
\end{align}
where $U\{1, 8\}$ denotes a discrete uniform distribution over the integers 1--8. Similarly, for non-elite drivers we set
\begin{align}
& \E \left[\mbox{Non-elite driver Finishing Position} \right]  = \E \left[ \operatorname{U}\{9, 20\} \right]=14.5,\label{eq2}
\end{align}
where $\operatorname{U}\{9, 29\}$ denotes a discrete uniform distribution over the integers 9--20. Elite drivers thus typically occupy top positions, with non-elite drivers finishing lower. The finishing rank of an elite-team driver is modeled as $N(4.5, \sigma_E^2)$. Similarly, the rank of a non-elite team driver is modeled as $N(14.5, \sigma_N^2)$. Thus, the probability that an elite-team driver wins a race can be calculated as
\begin{equation} \label{eq3}
\Pr(\mbox{Win})= \Pr(\mbox{Rank}\leq1.5)=\Phi\left(\frac{1.5-4.5}{\sigma_E}\right)=\Phi\left(-\frac{3}{\sigma_E}\right)
\end{equation}
where $\Phi(\cdot)$ denotes the $N(0,1)$ Cumulative Distribution Function (CDF). If one of the eight elite-team driver is guaranteed to win a race, consistent with trends from the past four completed seasons, it follows that
\begin{eqnarray}
    8\Phi\left(-\frac{3}{\sigma_E}\right)=1; \,\, 
    \sigma_E=-\frac{3}{\Phi^{-1}\left(\frac{1}{8}\right)}=2.607903.\label{eq4}
\end{eqnarray}
Similarly, the probability that a non-elite team driver finishes in the top 9 positions is given by
\begin{equation} \label{eq5}
    \Pr(\mbox{Rank}\leq{9.5})=\Phi\left(\frac{9.5-14.5}{\sigma_N}\right)=\Phi\left(-\frac{5}{\sigma_N}\right).
\end{equation}
Since there are only 8 elite-team drivers, at least one of the 12 non-elite drivers is guaranteed to finish in the top 9.  
It follows that
\begin{eqnarray}
    12\Phi\left(-\frac{5}{\sigma_N}\right)=1; \,\, \sigma_N= -\frac{5}{\Phi^{-1}\left(\frac{1}{12}\right)}=3.615344.\label{eq6}
\end{eqnarray}
Equation~\eqref{eq6} highlights that caution is needed when comparing drivers from different teams as the performance of non-elite drivers is inherently more variable with $\sigma_N > \sigma_E$\footnote{Equation~\eqref{eq6} can alternatively be derived by making the simplifying assumption that at least one of the 12 non-elite drivers is guaranteed to finish last; however, this assumption does not always hold in practice due to potential car failures or retirements.}.

Continuing in this way Table~\ref{tab:probabilities} presents analytical expressions for the probabilities of finishing in points scoring positions -- quantities that may be of particular interest to teams beyond winning \citep{MouraoBook}. The current Formula~1 scoring rules for the 2025 season are given in Table~\ref{tab:points}. Whilst some analytical results can be obtained from Tables~\ref{tab:probabilities}--\ref{tab:points}, the analysis can be simplified using Monte Carlo simulation. As an illustration, a full Formula~1 season for elite-team drivers can be simulated using 
\begin{enumerate}[label=(\roman*)] 
    \item Simulate a finishing position from $N(4.5, \sigma^2_E)$.
    \item Round to the nearest integer between 1 and 20.
    \item Allocate points according to Table~\ref{tab:points}.
    \item Repeat Steps (i)--(iii) for the rest of the races in a season.
\end{enumerate}

%\begin{table}[t!]
%\centering
%\small
%\renewcommand{\arraystretch}{1.15}
%\setlength{\tabcolsep}{8pt}
%\begin{tabular}{l>{\columncolor{gray!10}}l>{\columncolor{white}}l}
%\toprule
%\textbf{Probability} & \textbf{Elite-Team Driver} & \textbf{Non-Elite Team Driver} \\
%\midrule
%$\Pr(\text{Win})$             & $\Phi\left(-\frac{3}{\sigma_E}\right)$  & $\Phi\left(-\frac{13}{\sigma_N}\right)$ \\
%$\Pr(\text{Position }k), \; k=2,\ldots,10$          & $\Phi\left(-%\frac{4-k}{\sigma_E}\right)-\Phi\left(-\frac{5-k}{\sigma_E}\right)$  & $\Phi\left(-\frac{14-k}{\sigma_N}\right)-\Phi\left(-\frac{15-k}{\sigma_N}\right)$ \\
%$\Pr(\text{Podium finish})$   & $\Phi\left(-\frac{1}{\sigma_E}\right)$  & $\Phi\left(-\frac{11}{\sigma_N}\right)$ \\
%$\Pr(\text{Top 8 finish})$    & $\Phi\left(\frac{4}{\sigma_E}\right)$   & $\Phi\left(-\frac{6}{\sigma_N}\right)$ \\
%$\Pr(\text{Top 10 finish})$   & $\Phi\left(\frac{6}{\sigma_E}\right)$   & $\Phi\left(-\frac{4}{\sigma_N}\right)$ \\
%\bottomrule
%\end{tabular}
%\caption{Analytical expressions for race outcome probabilities derived from the univariate normal model (Section~\ref{sec:univariate}), where $\Phi$ is the CDF of a standard normal and $\sigma_E$, $\sigma_N$ denote the standard deviation of elite and non-elite team finishing positions (Equations~\eqref{eq4} and \eqref{eq6}).}
%\label{tab:probabilities}
%\end{table}

\begin{table}[t!]
\centering
\footnotesize
\renewcommand{\arraystretch}{1.15}
\setlength{\tabcolsep}{8pt}
\begin{tabular}{l>{\columncolor{gray!10}}l>{\columncolor{white}}l}
\toprule
\textbf{Probability} & \textbf{Elite-Team Driver} & \textbf{Non-Elite Team Driver} \\
\midrule
$\Pr(\text{Win})$             & $\Phi\left(-\frac{3}{\sigma_E}\right)$  & $\Phi\left(-\frac{13}{\sigma_N}\right)$ \\
$\Pr(\text{Second})$          & $\Phi\left(-\frac{2}{\sigma_E}\right)-\Phi\left(-\frac{3}{\sigma_E}\right)$  & $\Phi\left(-\frac{12}{\sigma_N}\right)-\Phi\left(-\frac{13}{\sigma_N}\right)$ \\
$\Pr(\text{Third})$           & $\Phi\left(-\frac{1}{\sigma_E}\right)-\Phi\left(-\frac{2}{\sigma_E}\right)$  & $\Phi\left(-\frac{11}{\sigma_N}\right)-\Phi\left(-\frac{12}{\sigma_N}\right)$ \\
$\Pr(\text{Fourth})$          & $\frac{1}{2}-\Phi\left(-\frac{1}{\sigma_E}\right)$  & $\Phi\left(-\frac{10}{\sigma_N}\right)-\Phi\left(-\frac{11}{\sigma_N}\right)$ \\
$\Pr(\text{Fifth})$           & $\Phi\left(\frac{1}{\sigma_E}\right)-\frac{1}{2}$  & $\Phi\left(-\frac{9}{\sigma_N}\right)-\Phi\left(-\frac{10}{\sigma_N}\right)$ \\
$\Pr(\text{Sixth})$           & $\Phi\left(\frac{2}{\sigma_E}\right)-\Phi\left(\frac{1}{\sigma_E}\right)$  & $\Phi\left(-\frac{8}{\sigma_N}\right)-\Phi\left(-\frac{9}{\sigma_N}\right)$ \\
$\Pr(\text{Seventh})$         & $\Phi\left(\frac{3}{\sigma_E}\right)-\Phi\left(\frac{2}{\sigma_E}\right)$  & $\Phi\left(-\frac{7}{\sigma_N}\right)-\Phi\left(-\frac{8}{\sigma_N}\right)$ \\
$\Pr(\text{Eighth})$          & $\Phi\left(\frac{4}{\sigma_E}\right)-\Phi\left(\frac{3}{\sigma_E}\right)$  & $\Phi\left(-\frac{6}{\sigma_N}\right)-\Phi\left(-\frac{7}{\sigma_N}\right)$ \\
$\Pr(\text{Ninth})$           & $\Phi\left(\frac{5}{\sigma_E}\right)-\Phi\left(\frac{4}{\sigma_E}\right)$  & $\Phi\left(-\frac{5}{\sigma_N}\right)-\Phi\left(-\frac{6}{\sigma_N}\right)$ \\
$\Pr(\text{Tenth})$           & $\Phi\left(\frac{6}{\sigma_E}\right)-\Phi\left(\frac{5}{\sigma_E}\right)$  & $\Phi\left(-\frac{4}{\sigma_N}\right)-\Phi\left(-\frac{5}{\sigma_N}\right)$ \\
\midrule
$\Pr(\text{Podium finish})$   & $\Phi\left(-\frac{1}{\sigma_E}\right)$  & $\Phi\left(-\frac{11}{\sigma_N}\right)$ \\
$\Pr(\text{Top 8 finish})$    & $\Phi\left(\frac{4}{\sigma_E}\right)$   & $\Phi\left(-\frac{6}{\sigma_N}\right)$ \\
$\Pr(\text{Top 10 finish})$   & $\Phi\left(\frac{6}{\sigma_E}\right)$   & $\Phi\left(-\frac{4}{\sigma_N}\right)$ \\
\bottomrule
\end{tabular}
\caption{Analytical expressions for race outcome probabilities derived from the univariate normal model (Section~\ref{sec:univariate}), where $\Phi$ is the CDF of a standard normal and $\sigma_E$, $\sigma_N$ denote the standard deviation of elite and non-elite team finishing positions (Equations~\eqref{eq4} and \eqref{eq6}).}
\label{tab:probabilities}
\end{table}

\begin{table}[t!]
\centering
\footnotesize
\setlength{\tabcolsep}{6pt}
\begin{tabular}{rrrr}
\toprule
\multicolumn{2}{c}{\textbf{Full Race}} & \multicolumn{2}{c}{\textbf{Sprint Race}} \\
\cmidrule(lr){1-2} \cmidrule(lr){3-4}
\textbf{Position} & \textbf{Points} & \textbf{Position} & \textbf{Points} \\
\midrule
1  & 25 & 1 & 8 \\
2  & 18 & 2 & 7 \\
3  & 15 & 3 & 6 \\
4  & 12 & 4 & 5 \\
5  & 10 & 5 & 4 \\
6  & 8  & 6 & 3 \\
7  & 6  & 7 & 2 \\
8  & 4  & 8 & 1 \\
9  & 2  &   &   \\
10 & 1  &   &   \\
\bottomrule
\end{tabular}
\caption{Points awarded to drivers in Formula~1 for Full Races (top~10) and Sprint Races (top~8) under the rules of the 2025 season; all other positions receive 0 points.}
\label{tab:points}
\end{table}

\section{Bivariate modelling of team performance} \label{sec:bivariate}
In this section, we extend the previous univariate model to a bivariate model for the performance of two drivers within the same team. This explicitly accounts for negative correlations in teammates' finishing positions. Following Section~\ref{sec:univariate} we model the finishing rank of an elite-team driver as ${N}(4.5, \sigma_E^2)$, where $\sigma_E^2$ is given by Equation~\eqref{eq4}. The joint performance of the two drivers is then represented by a bivariate normal distribution:
\begin{eqnarray}
    \left(\begin{array}{l}
         r_1  \\
         r_2 
    \end{array}\right) \sim N \left(\left(\begin{array}{l}
         4.5  \\
         4.5 
    \end{array}\right),\left(\begin{array}{ll}
       \sigma^2_E  & \sigma_{EE} \\
    \sigma_{EE}     & \sigma^2_E
    \end{array}\right)\right),\label{eq7}
\end{eqnarray}
where $r_i$ denotes the rank of Driver $i$, and $\sigma_{EE}$ is the  covariance. The value of $\sigma_{EE}$ in equation~\eqref{eq7} can be derived as follows. Using the linear transformation property of the normal distribution, \citep{BinghamFry}, the sum of the ranks satisfies $r_1+r_2 \sim {N}(9, 2\sigma^2_E+2\sigma_{EE})$. The sum of the ranks must satisfy $\Pr(r_1 + r_2 \leq 3) = 0$, giving 
$$
\Pr(r_1 + r_2 \leq 3) = \Phi \left( \frac{3-9}{\sqrt{2\sigma^2_E+2\sigma_{EE}}}    \right) = 0 .
$$
Tabulated values of the standard normal from \cite{Neave} give $\Phi(-4.9) = 0$. Thus,
\begin{equation}
\frac{3-9}{\sqrt{2\sigma^2_E+2\sigma_{EE}}} = -4.9 ; \,\,
\sigma_{EE}  = \frac{36}{48.02} - \sigma_E^2 =  -6.051472. \label{eq8}
\end{equation}
For a non-elite team equation~\eqref{eq7} is replaced by
\begin{eqnarray}
    \left(\begin{array}{l}
         r_1  \\
         r_2 
    \end{array}\right)\sim N \left(\left(\begin{array}{l}
         14.5  \\
         14.5 
    \end{array}\right),\left(\begin{array}{ll}
       \sigma^2_N  & \sigma_{NN} \\
    \sigma_{NN}     & \sigma^2_N
    \end{array}\right)\right). \label{eq9}
\end{eqnarray}
The value of the covariance $\sigma_{NN}$ in equation~\eqref{eq9} can be derived as follows. From the linear transformation property of the normal distribution $r_1+r_2 \sim N(29, 2\sigma^2_N+2\sigma_{NN})$. The sum of the ranks must satisfy $\Pr(r_1+r_2\leq39)=1$.
%, yielding
%$$
%\Pr(r_1 + r_2 \leq39) = \Phi \left( \frac{39-29}{\sqrt{2\sigma^2_N+2\sigma_{NN}}}    \right) = 1 .
%$$
Tabulated values in \cite{Neave} give $\Phi(4.9)=1$, so 
\begin{equation}
\frac{39-29}{\sqrt{2\sigma^2_N+2\sigma_{NN}}} = 4.9 ; \,\,
\sigma_{NN} = \frac{100}{48.02} - \sigma_{N}^2 = -10.98825. \label{eq10}
\end{equation}
As in Section~\ref{sec:univariate}, the analysis of finishing ranks can be performed via Monte Carlo simulation. As an illustration, results for negatively correlated drivers within the same elite-team can be simulated using
\begin{enumerate}
    \item Simulate finishing positions from ${N}\left(\left(\begin{array}{l}
         4.5 \\
         4.5
    \end{array}\right), \left(\begin{array}{ll}
  \sigma^2_E       & \sigma_{EE} \\
     \sigma_{EE}    & \sigma^2_E
    \end{array}\right)\right)$.
    \item Round to the nearest integers between 1 and 20.
    \item Allocate points according to Table~\ref{tab:points}.
    \item Repeat Steps 1--3 for the rest of the races in a season.
\end{enumerate}

\section{Empirical applications} \label{sec:empirical}

In this section, we compare results models in Section~\ref{sec:univariate}--\ref{sec:bivariate} with the last fully completed 2025 season, which featured 24 Full Races and 6 Sprint Races. Table~\ref{tab:montecarlo} presents Monte Carlo simulation results for the expected value and 95\% confidence intervals (CIs) for season-long points totals across elite and non-elite drivers and teams. These CIs provide a benchmark for  reasonable performance expectations, across all sectors, and can then be compared with actual performances from the 2025 season.

\begin{table}[t!]
\centering
\footnotesize
\renewcommand{\arraystretch}{1.05} % slightly taller rows
\setlength{\tabcolsep}{9pt}       % spacing between columns
\begin{tabular}{l r r}
\toprule
\textbf{Category} & \textbf{Mean points} & \textbf{95\% CI} \\
\midrule
\rowcolor{gray!10} Elite driver & 315.456 & (253-381) \\
\rowcolor{gray!10} Elite team   & 630.91 & (594--669) \\
Non-elite driver                & 10.636       & (1--29) \\
Non-elite team                  & 21.305      & (5--44) \\
\bottomrule
\end{tabular}
\caption{Monte Carlo simulation of total season points under the 2025 Formula~1 scoring system (Table~\ref{tab:points}), based on 1,000,000 simulations. Drivers are simulated with the univariate model (Section~\ref{sec:univariate}) and teams with the bivariate model (Section~\ref{sec:bivariate}). Reported values are mean total points and 95\% CIs for elite and non-elite drivers and teams.}
\label{tab:montecarlo}
\end{table}

Results for individual drivers are reported in Table~\ref{tab:montecarlo2}. Several elite drivers exceed expectations (Norris, Piastri, and Verstappen). Some elite drivers may be under-performing (Antonelli, Hamilton, Leclerc and Tsunoda). Remarkably, most non-elite drivers exceed expectations, highlighting the talent and depth of competition in Formula~1. Fernando Alonso is a particularly highly rated driver \citep{FryF1} who outperforms the benchmark. Colapinto is the only non-elite driver potentially under-performing. Results in Table~\ref{tab:montecarlo3} provide further insights into team performance. McLaren, exceed expectations after an outstanding season. In contrast, the remaining elite teams seem to under-perform. Most non-elite teams, exceed expectations. In contrast, Alpine perform in line with reasonable performance expectations.

\begin{table}[t!]
\centering
\footnotesize
\renewcommand{\arraystretch}{1.05} % slightly taller rows
\setlength{\tabcolsep}{7pt}       % spacing between columns
\begin{tabular}{l l r c}
\toprule
\textbf{Driver} & \textbf{Team} & \multicolumn{1}{c}{\textbf{Points}} & \textbf{Performance} \\
\midrule

% Elite-team drivers
\rowcolor{gray!10} Lando Norris        & McLaren       & 423 & ↑ \\
\rowcolor{gray!10} Oscar Piastri       & McLaren       & 410 & ↑ \\
\rowcolor{gray!10} George Russell      & Mercedes      & 319 & $\rightarrow$ \\
\rowcolor{gray!10} Kimi Antonelli      & Mercedes      & 150 & ↓ \\
\rowcolor{gray!10} Max Verstappen      & Red Bull      & 421 & ↑ \\
\rowcolor{gray!10} Yuki Tsunoda        & Red Bull      & 33  & ↓ \\
\rowcolor{gray!10} Charles Leclerc     & Ferrari       & 242 & ↓ \\
\rowcolor{gray!10} Lewis Hamilton      & Ferrari       & 156 & ↓ \\

\addlinespace[2pt] % separation between elite and non-elite

% Non-elite drivers
Alexander Albon       & Williams      & 73  & $\uparrow$ \\
Carlos Sainz Jr       & Williams      & 64  & $\uparrow$ \\
Isack Hadjar          & Racing Bulls  & 51  & $\uparrow$ \\
Liam Lawson           & Racing Bulls  & 38  & $\uparrow$ \\
Fernando Alonso       & Aston Martin  & 56  & $\uparrow$ \\
Lance Stroll          & Aston Martin  & 33  & $\rightarrow$ \\
Oliver Bearman        & Haas          & 41  & $\uparrow$ \\
Esteban Ocon          & Haas          & 38  & $\uparrow$ \\
Nico H\"{u}lkenberg   & Sauber        & 51  & $\uparrow$ \\
Gabriel Bortoleto     & Sauber        & 19  & $\rightarrow$ \\
Pierre Gasly          & Alpine        & 22  & $\rightarrow$ \\
Franco Colapinto      & Alpine        & 0   & $\downarrow$ \\

\bottomrule
\end{tabular}
\caption{Total points scored by drivers in the 2025 season (elite-team drivers are highlighted). The \emph{Performance} column compares points with Monte Carlo CIs (Table~\ref{tab:montecarlo}), which provide benchmark expectations based on the univariate model (Section~\ref{sec:univariate}); Symbols indicate performance relative to expectations: $\uparrow$ = above, $\rightarrow$ = meeting, $\downarrow$ = below.}
\label{tab:montecarlo2}
\end{table}

\begin{table}[t!]
\centering
\footnotesize
\renewcommand{\arraystretch}{1.05}
\setlength{\tabcolsep}{7pt}
\begin{tabular}{l r c}
\toprule
\textbf{Team} & \multicolumn{1}{r}{\textbf{Points}} & \textbf{Performance} \\
\midrule
% Elite teams
\rowcolor{gray!10} McLaren      & 833 & $\uparrow$ \\
\rowcolor{gray!10} Mercedes     & 469 & $\downarrow$ \\
\rowcolor{gray!10} Red Bull     & 454 & $\downarrow$ \\
\rowcolor{gray!10} Ferrari      & 398 & $\downarrow$ \\
\addlinespace[2pt]
% Non-elite teams
Williams        & 137 & $\uparrow$ \\
Racing Bulls    & 92  & $\uparrow$ \\
Aston Martin    & 89  & $\uparrow$ \\
Haas            & 79  & $\uparrow$ \\
Sauber          & 70  & $\uparrow$ \\
Alpine          & 22  & $\rightarrow$ \\
\bottomrule
\end{tabular}
\caption{Total points scored by teams in the 2025 season (elite teams are highlighted). The \emph{Performance} column compares points with Monte Carlo CIs (Table~\ref{tab:montecarlo}), which provide benchmark expectations based on the bivariate model (Section~\ref{sec:bivariate}); Symbols indicate performance relative to expectations: $\uparrow$ = above, $\rightarrow$ = meeting, $\downarrow$ = below.}
\label{tab:montecarlo3}
\end{table}

\section{Conclusions} \label{sec:conclusion}

This paper quantifies reasonable performance expectations in sports \citep{FryJBR2021}, linking to related questions of competitive balance \citep{Plumley} and performance efficiency \cite{Rossi}. It also contributes to a wider literature disentangling driver and car-level effects in Formula~1 \citep{Bell,vankesteren,FryF1}. 

Formula~1 consists of teams of two drivers, with four elite teams -- Red Bull, Mercedes, Ferrari, and McLaren -- dominating the sport in recent seasons. Despite this, competition remains intense. In this paper, we develop univariate and bivariate normal models to assess reasonable performance expectations for drivers and teams. By explicitly incorporating the elite/non-elite distinction and within-team correlations, the models provide a more realistic evaluation of performance. We derive analytical results for winning probabilities, and conduct Monte Carlo simulations. These simulations provide benchmarks for drivers and teams, enabling a rigorous evaluation of performance in the last fully completed 2025 season. We can thus who exceeded, met, or fell short of expectations. However, Formula~1 remains highly competitive and influenced by randomness, so individual race results should be interpreted cautiously. In some cases, performance may need to be revised downward, as discussed in Appendix~\ref{appendix}.

Empirical results obtained contribute to the wider literature \citep{Bell, vankesteren, FryF1}. At the driver level, Norris, Piastri, and Verstappen exceeded expectations. Several elite drivers under-perform relative to model predictions. At the team level, McLaren exceeded expectations, whilst other elite teams Mercedes, Red Bull and Ferrari under-perform. Remarkably, most non-elite drivers exceed expectations, with Colapinto the only exception. This emphasizes the increasingly competitive landscape beyond the elite top teams.

Overall, the results highlight the value of incorporating both the elite/non-elite distinction and within-team correlations when benchmarking performance, offering a more nuanced and quantitative understanding of outcomes in Formula~1. Potential applications include driver selection and contract negotiations. However, there remains an enduring need to interpret the numbers with care. Appendix~\ref{appendix} discusses cases where performance benchmarks may need to be revised downward, particularly in the presence of a dominant manufacturer or when evaluating rookie drivers. At the same time, in highly competitive sporting environments, technological advantages often dissipate quickly \citep{Rockerbie}, implying that the returns to additional wind tunnel and CFD investment are likely to remain uncertain.

\appendix

\begin{table}[t!]
\centering
\footnotesize
\renewcommand{\arraystretch}{1.05} % slightly taller rows
\setlength{\tabcolsep}{9pt}       % spacing between columns
\begin{tabular}{l r r}
\toprule
\textbf{Category} & \textbf{Mean points} & \textbf{95\% CI} \\
\midrule
\rowcolor{gray!10} Elite driver & 195.871 & (147-249) \\
\rowcolor{gray!10} Elite team   & 391.733 & (361--424) \\
\bottomrule
\end{tabular}
\caption{Monte Carlo simulation of total season points under the 2025 Formula~1 scoring system (Table~\ref{tab:points}), based on 1,000,000 simulations. Drivers are simulated with the univariate model (Section~\ref{sec:univariate}) and teams with the bivariate model (Section~\ref{sec:bivariate}), with parameter $\mu_E=5.5$. Reported values are mean total points and 95\% CIs for elite drivers and teams.}
\label{tab:montecarlo2}
\end{table}

\section{Further performance benchmarking considerations} \label{appendix}

\paragraph{Revised performance targets in the face of a single dominant manufacturer.} Race car performance is known to depend on technological factors such as aerodynamics \citep{Katz}, tyre degradation \citep{West}, and engine power \citep{Boretti}. McLaren's dominance in the 2025 season is is attributed to their ability to maintain optimal tyre temperature. This raises the question of how performance benchmarks should be adjusted in the presence of a clear technological advantage at the manufacturer level. 

To account for a single dominant manufacturer, equation \eqref{eq1} should be modified as follows:
\begin{align}
& \E\left[\mbox{Elite driver Finishing Position}\right]  = \E \left[ \operatorname{U}\{3, 8\} \right] = 5.5. \label{eq11}
\end{align}
Applying this adjustment to the mean finishing position $\mu_E$ of elite-team drivers yields the simulation results and revised performance benchmarks reported in Table~\ref{tab:montecarlo2}.

\paragraph{Performance benchmarking for rookie drivers.} Suppose a rookie driver joins an elite team but requires one year to ``bed-in'' before fully adjusting and reaching reasonable performance expectations in year two. In this case, performance in the first year would only need to be sufficient to suggest that the targets in Table~\ref{tab:montecarlo} are likely to be met by the end of the second year. Under this assumption, halving the original performance targets yields an expected points total of 157.73 points, with a 95\% confidence interval of 126.5--190.5 points.

\section*{Funding}
No funding was received.

\section*{Disclosure of interest}
The authors declare no conflicting interests.

\section*{Acknowledgements}
The authors gratefully acknowledge the helpful and constructive comments of two anonymous referees and participants at the \textit{Active Territories: Sport and Leisure} conference, \'{e}klore-ed School of Management, Pau, France, May 2026. The usual disclaimer applies.

\bibliographystyle{elsarticle-harv}   % Bibliography style for Elsevier, e.g. Economics Letters
\bibliography{bib_F1_Benchmark.bib}

@article{Bell,
	author = {Bell, A. and Smith, J. and Sabel, C. E. and Jones, K.},
	journal = {J. Quant. Anal. Sports},
	pages = {99--112},
	title = {Formula for success: {M}ultilevel modelling of {F}ormula {O}ne driver and constructor performance, 1950-2014},
	volume = {12},
	year = {2016}}

@article{Boretti,
	author = {Boretti, A.},
	journal = {Nonlinear Eng.},
	pages = {28--34},
	title = {Energy flow of a 2018 {F}{I}{A} {F}1 racing car and proposed changes to the powertrain rules},
	volume = {9},
	year = {2020}}

@article{FryJBR2021,
	author = {Fry, J. and Serbera, J.-P. and Wilson, R.},
	journal = {J. of Bus. Res.},
	pages = {445--453},
	title = {Managing performance expectations in association football},
	volume = {135},
	year = {2021}}

@article{FryF1,
	author = {Fry, J. and Brighton, T. and Fanzon, S.},
	journal = {Econ. Lett.},
	pages = {111671},
	title = {Faster Identification of Faster {F}ormula 1 Drivers via Time-Rank Duality},
	volume = {237},
	year = {2024}}

@article{Katz,
	author = {Katz, J.},
	journal = {Annu. Rev. Fluid Mech.},
	pages = {27-63},
	title = {Aerodynamics of race cars},
	volume = {38},
	year = {2006}}

@book{Neave,
	author = {Neave, H. R.},
	publisher = {George Allen and Unwin},
	title = {Statistics Tables for Mathematicians, Engineers, Economists and the Behavioural and Management Sciences},
	year = {1978}}

@article{Plumley,
	author = {Plumley, D. and Ramchandani, G. and Wilson, R.},
	journal = {Sport Bus. and Manag.},
	pages = {118-133},
	title = {The unintended consequences of financial fair play: An examination of competitive balance across five European football leagues},
	volume = {9},
	year = {2019}}

@article{Rockerbie,
	author = {Rockerbie, D. W. and Easton, S. T.},
	journal = {Appl. Econ.},
	pages = {6272-6285},
	title = {Race to the podium: separating and conjoining the car and driver in {F}1 racing},
	volume = {54},
	year = {2022}}

@article{Rossi,
	author = {Rossi, G. and Goossens, D. and Di Tanna, G. L. and Addessa, F.},
	journal = {Eur. Sport Manag. Q.},
	pages = {583-604},
	title = {Football team performance efficiency and effectiveness in a corruptive context: the {C}alciopoli case},
	volume = {19},
	year = {2019}}

@article{Scarf2,
	author = {Scarf, P. and Khare, A. and Alotaibi, N.},
	journal = {IMA J. of Manag. Math.},
	pages = {53--73},
	title = {On Skill and Chance in Sport},
	volume = {33},
	year = {2022}}

@article{Scarf,
	author = {Scarf, P. and Parama, R. and McHale, I.},
	journal = {Eur. J. of Oper. Res.},
	pages = {721--730},
	title = {On Outcome Uncertainty and Scoring Rates in Sport: The Case of International {R}ugby {U}nion},
	volume = {273},
	year = {2019}}

@article{vankesteren,
	author = {Van Kesteren, E-J. and Bergkamp, T.},
	journal = {J. Quant. Anal. Sports},
	pages = {273-293},
	title = {Baysian analysis of Formula One race results: Disentangling driver skill and constructor advantage},
	volume = {19},
	year = {2023}}

@article{Wesselbaum,
	author = {Wesselbaum, D. and Owen, P. D.},
	journal = {Aust. Econ. Rev.},
	pages = {164--173},
	title = {The value of pole position in {F}ormula 1 history},
	volume = {54},
	year = {2021}}

@article{West,
	author = {West, W. J. and Limebeer, D. J. N.},
	journal = {Veh. Syst. Dyn.},
	pages = {1--19},
	title = {Optimal tyre management for a high-performance race car},
	volume = {60},
	year = {2022}}

@book{BinghamFry,
	author = {Bingham, N. H. and Fry, J. M.},
	publisher = {Springer},
	title = {Regression: Linear models in statistics},
	year = {2010}}

@book{MouraoBook,
	author = {Mour\~ao, P. and Macedo, A. and Oliveira, A.},
	publisher = {Springer},
	title = {The {E}conomics of {M}oto{GP}: The {C}osts, {F}inancing and {C}ompetitive {B}alance of a {R}ising {M}otorsport},
	year = {2025}}

 \end{document}